\documentclass[conference]{IEEEtran}

\usepackage{graphicx}
\usepackage{booktabs}
\usepackage{siunitx} 
\usepackage{amsmath,amssymb,amsfonts}
\usepackage{array}
\usepackage{tabularx}
\usepackage{enumitem}
\usepackage{makecell}
\usepackage{listings}
\usepackage{xcolor}
\usepackage{algorithm}
\usepackage{algpseudocode}
\usepackage{tikz}
\usepackage{pgfplots}
\pgfplotsset{compat=1.18}

\usepackage[caption=false,font=normalsize,labelfont=sf,textfont=sf]{subfig}

\usepackage{adjustbox}

\usepackage{balance}

\usepackage[colorlinks=true,linkcolor=black,citecolor=black,urlcolor=black]{hyperref}

\definecolor{codegreen}{rgb}{0,0.6,0}
\definecolor{codegray}{rgb}{0.5,0.5,0.5}
\definecolor{codepurple}{rgb}{0.58,0,0.82}
\definecolor{backcolour}{rgb}{0.95,0.95,0.92}
\definecolor{codered}{rgb}{0.91,0.58,0.53}

\definecolor{softblue}{RGB}{102,178,255}
\definecolor{softred}{RGB}{255,153,153}
\definecolor{softgreen}{RGB}{153,255,153}
\definecolor{softyellow}{RGB}{255,255,153}
\definecolor{softcyan}{RGB}{153,255,255}
\definecolor{softmagenta}{RGB}{255,153,255}
\definecolor{softorange}{RGB}{255,200,150}

\def\BibTeX{{\rm B\kern-.05em{\sc i\kern-.025em b}\kern-.08em
    T\kern-.1667em\lower.7ex\hbox{E}\kern-.125emX}}

\lstdefinestyle{mystyle}{
    backgroundcolor=\color{backcolour},   
    commentstyle=\color{black},
    keywordstyle=\color{black},
    numberstyle=\scriptsize\color{codegray},
    stringstyle=\color{black},
    basicstyle=\ttfamily\scriptsize, 
    breakatwhitespace=false,         
    breaklines=true,                 
    captionpos=b,                    
    keepspaces=true,                 
    numbers=left,                    
    numbersep=5pt,                  
    showspaces=false,                
    showstringspaces=false,
    showtabs=false,                  
    tabsize=2
}

\lstdefinestyle{mystyle_1}{
    backgroundcolor=\color{codered},   
    commentstyle=\color{codegreen},
    keywordstyle=\color{black},
    numberstyle=\scriptsize\color{codered},
    stringstyle=\color{codered},
    basicstyle=\ttfamily\scriptsize, 
    breakatwhitespace=false,         
    breaklines=true,                 
    captionpos=b,                    
    keepspaces=true,                 
    numbers=left,                    
    numbersep=5pt,                  
    showspaces=false,                
    showstringspaces=false,
    showtabs=false,                  
    tabsize=2
}

\newcommand{\tool}{\textsc{Xamt}}

\begin{document}

\title{\tool: Cross-Framework API Matching for Testing Deep Learning Libraries}

\author{
Bin Duan, 
Ruican Dong, 
Naipeng Dong, 
Dan Dongseong Kim, 
Guowei Yang* \\
School of Electrical Engineering and Computer Science, \\
The University of Queensland, Brisbane, Australia \\
b.duan@uq.edu.au, ruican.dong@uq.net.au, n.dong@uq.edu.au, dan.kim@uq.edu.au, guowei.yang@uq.edu.au
\thanks{*Corresponding author: guowei.yang@uq.edu.au}
}

\maketitle

\begingroup
\renewcommand\thefootnote{*}
\renewcommand{\footnoterule}{}
\footnotetext{Corresponding author.}
\endgroup

\begin{abstract}

Deep learning powers critical applications such as autonomous driving, healthcare, and finance, where the correctness of underlying libraries is essential. Bugs in widely used deep learning APIs can propagate to downstream systems, causing serious consequences.
While existing fuzzing techniques detect bugs through intra-framework testing across hardware backends (CPU vs. GPU), they may miss bugs that manifest identically across backends and thus escape detection under these strategies.
To address this problem, we propose \tool, a cross-framework fuzzing method that tests deep learning libraries by matching and comparing functionally equivalent APIs across different frameworks. \tool\ matches APIs using similarity-based rules based on names, descriptions, and parameter structures. It then aligns inputs and applies variance-guided differential testing to detect bugs.
We evaluated \tool\ on five popular frameworks, including PyTorch, TensorFlow, Keras, Chainer, and JAX.
\tool\ matched 839 APIs and identified 238 matched API groups, and detected 17 bugs, 12 of which have been confirmed.
Our results show that \tool\ uncovers bugs undetectable by intra-framework testing, especially those that manifest consistently across backends. \tool\ offers a complementary approach to existing methods and offers a new perspective on the testing of deep learning libraries.


\end{abstract}
\begin{IEEEkeywords}
Fuzzing, Deep Learning Libraries
\end{IEEEkeywords}

\section{Introduction}

Deep learning has become integral to real-world systems in domains such as autonomous driving~\cite{rao2018deep, bogdoll2022anomaly}, healthcare~\cite{miotto2018deep, arabahmadi2022deep}, and finance~\cite{heaton2017deep, venkateswarlu2022efficient}. These applications depend heavily on the correctness of deep learning libraries, whose APIs are widely used for model construction, training, and inference~\cite{singh2020explainable}. Considering their widespread use and impact on downstream applications, implementation bugs in deep learning library APIs, such as incorrect outputs, unstable gradients, or silent logic flaws, can propagate into model behavior and lead to significant errors in real-world systems.

To detect such bugs, recent efforts apply fuzzing techniques~\cite{li2018fuzzing, xie2022docter}, which aim to uncover edge-case errors by generating diverse and often unexpected inputs, and applying oracle-based reasoning to detect anomalies. For deep learning libraries, fuzzing techniques are typically divided into model-level fuzzing and API-level fuzzing.
Model-level fuzzing~\cite{guo2020audee, gu2022muffin, wang2020deep} focuses on generating or mutating entire DNN models and observing their behavior under various inputs.
In contrast, API-level fuzzing~\cite{wei2022free, deng2023largetitan} directly targets individual APIs within deep learning libraries, systematically exploring their parameter space to reveal abnormal behaviors. These approaches have proven effective in exposing a variety of bugs.

However, despite their successes, most existing approaches share a fundamental limitation: their oracles mainly rely on intra-framework differential testing across backends (CPU vs. GPU). The key insight of these strategies lies that discrepancies between different backend outputs indicate potential bugs of the tested APIs. 
However, not all bugs manifest as output differences across different backends. To detect bugs that persist even when CPU and GPU outputs appear identical, relying solely on backend comparisons is insufficient. 

To address this problem, 
this paper proposes \tool, a cross-framework fuzzing method that matches functionally equivalent APIs from different deep learning libraries and compares their behavior to detect bugs. \tool\ begins by matching API groups using similarity-based matching rules derived from official documentation, considering API names, natural language descriptions, and parameter structures. Once matched, it aligns the parameter interfaces of each group and generates valid test inputs that respect the expected types and shapes. Then, a variance-guided differential fuzzing strategy is applied, which iteratively mutates these inputs to explore and amplify the output divergence. Discrepancies are detected using three oracles: crash, NaN, and output inconsistency across frameworks. 
Compared to existing intra-framework methods, \tool\ can detect bugs that manifest identically on CPU vs GPU. 


\begin{figure*}[t!]
  \centering
  \includegraphics[width=1\linewidth]{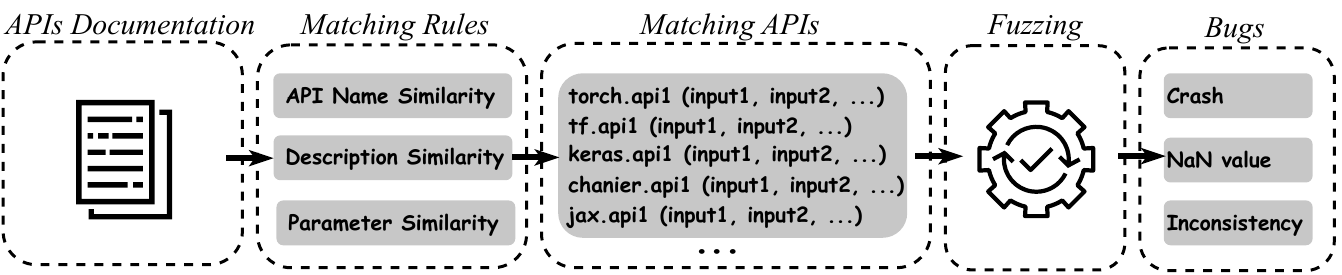}
  \caption{Overview of \tool.}
  \label{fig:overview}
\end{figure*}


We evaluated \tool\ on five widely used deep learning frameworks: PyTorch~\cite{paszke2019pytorch}, TensorFlow~\cite{abadi2016tensorflow}, Keras~\cite{chollet2015keras}, Chainer~\cite{tokui2015chainer}, and JAX~\cite{jax2018github}. In total, \tool\ matched 839 APIs, resulting in 238 matched API groups. \tool\ achieved good cross-framework compatibility, and detected 17 bugs from five frameworks, of which 12 have been confirmed. These bugs exhibit consistent behaviors between CPU and GPU, making them undetectable by existing intra-framework testing methods.

In summary, this paper makes the following contributions: 
\begin{itemize}[leftmargin=*]
\item \textbf{Idea}. 
We introduce \tool, a cross-framework fuzzing approach that matches and compares the behavior of functionally equivalent APIs.
\tool\ identifies functionally equivalent APIs from official documentation and applies variance-guided differential testing to detect potential bugs.

\item \textbf{Tool}. 
We implement \tool\ as a modular testing framework that includes API matching, parameter alignment, input generation, and differential fuzzing. \tool\ extracts API names, descriptions, and signatures from official documentation to form equivalence groups, generates valid inputs, and applies variance-guided mutation with crash, NaN, and inconsistency oracles. All code is publicly available. 

\item \textbf{Evaluation}. 
We evaluate \tool\ on five popular deep learning frameworks. \tool\ matched 839 APIs and identified 238 functionally equivalent API groups, and detected 17 bugs, 12 of which have been confirmed.

\end{itemize}

\section{Approach}

Fig.~\ref{fig:overview} shows an overview of \tool. We begin by collecting official API documentation from five mainstream deep learning frameworks and extracting structured API information, including API names, descriptions, and parameters. Based on this, in Section~\ref{sec:3.1}, we propose three similarity-based matching rules: API name, description, and parameter similarity. In Section~\ref{sec:3.2}, based on the three rules defined in the previous section, we match functionally equivalent APIs across different frameworks and group these equivalent APIs together. Subsequently, we align the input parameters for each API group to ensure that all frameworks receive consistent and valid inputs. Next, in Section~\ref{sec:3.3}, \tool\ generates consistent input seeds and invokes all matched APIs. To uncover behavioral inconsistencies, \tool\ applies variance-guided differential testing, which iteratively mutates valid inputs to maximize output divergence across frameworks. Outputs are evaluated using three oracles, crash, NaN, and inconsistency, to detect abnormal behaviors in Section~\ref{sec:3.4}.

\subsection{Defining Similarity-Based Matching Rules}
\label{sec:3.1}

\begin{figure*}[ht]
  \centering
  \includegraphics[width=1\linewidth]{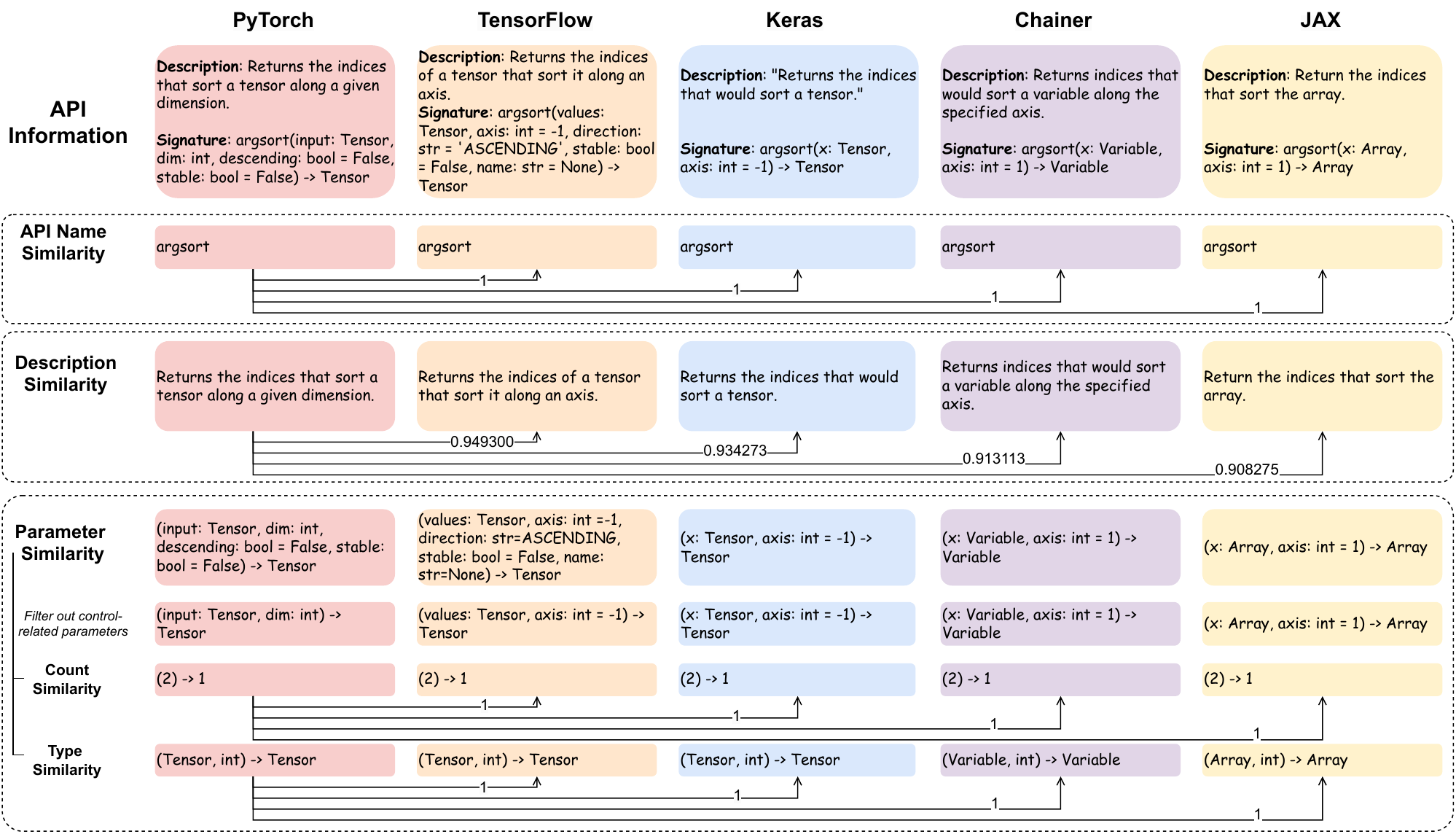}
  \caption{Matching Rules.}
  \label{fig:matchingrules}
\end{figure*}

To improve the reliability of API matching for functionally equivalent APIs across frameworks, we define three similarity-based matching rules to quantify the degree of correspondence between APIs: \textit{API Name Similarity}, \textit{Description Similarity}, and \textit{Parameter Similarity}, as shown in Fig~\ref{fig:matchingrules}. Each of these metrics captures a different aspect of the API. API names provide intuitive initial matching clues due to standard naming conventions in frameworks; descriptions capture functional semantics that might not be reflected directly by names; and parameter structures ensure precise structural alignment at the implementation level. By measuring API similarity from these three perspectives, we aim to accurately match functionally equivalent APIs across different frameworks.

\subsubsection{API Name Similarity}
This similarity measures the lexical similarity between API names in frameworks. API names often reflect their core functionality, but naming conventions can differ significantly across libraries. Therefore, we normalize API names by converting them to lowercase and removing framework-specific namespace prefixes. We then compute their normalized edit distance to define the similarity score:

\begin{equation}
{SIM_{APIName}} = 1 - \frac{d(a, b)}{\max({len}(a), {len}(b))}
\label{eq:api-name-sim}
\end{equation}

where $a$ and $b$ are the normalized API names, and $d(a, b)$ is the Levenshtein distance~\cite{levenshtein1966binary}, defined as the minimum number of single-character edits (insertions, deletions, or substitutions) required to transform $a$ into $b$. A score closer to 1 indicates higher lexical similarity, which often correlates with conceptual consistency across frameworks.

As shown in the Figure~\ref{fig:matchingrules}, all five frameworks use the identical function name \texttt{\small argsort}. Accordingly, the similarity score between every pair is 1. This case demonstrates that even across diverse libraries, consistent naming can offer a reliable cue for functional equivalence.

\subsubsection{Description Similarity}

This similarity estimates the semantic equivalence of APIs by comparing their descriptions in natural language, which are typically extracted from official documentation and reflect the intended behavior of each API.
To assess the semantic similarity of API descriptions, we adopt Sentence-BERT~\cite{reimers2019sentence}, a Transformer-based model that encodes sentences into dense embeddings while preserving contextual semantics. Each API description is converted into a high-dimensional vector:

\begin{equation}
{Vector}_{x} = \text{Sentence-BERT}({Description}_x)
\label{eq:SS_vector}
\end{equation}

Semantically similar sentences produce embeddings that are closely positioned in the vector space. Therefore, the cosine similarity between these vectors provides an effective measure of semantic equivalence between API descriptions:

\begin{equation}
SIM_{Des} = \frac{{Vector}_{a} \cdot {Vector}_{b}}{\|{Vector}_{a}\| \, \|{Vector}_{b}\|}
\label{eq:DS_cosine}
\end{equation}

where ${Vector_a}$ and ${Vector_b}$ are the vector representations of the API descriptions from different frameworks.
A higher similarity score implies a closer semantic relationship, suggesting that the APIs may provide similar or related functionalities.
As illustrated in Figure~\ref{fig:matchingrules}, the \texttt{\small argsort} API has functionally consistent descriptions across frameworks. Although phrasing differs, all descriptions express the same core functionality, returning the indices that would sort the input tensor along a specified axis. Sentence-BERT embeddings yield high cosine similarity scores (ranging from 0.908 to 0.949), confirming strong semantic alignment across implementations.

\subsubsection{Parameter Similarity}

This similarity quantifies the structural alignment between APIs by comparing their functional input parameters. Functionally equivalent APIs typically share similar parameter interfaces. However, practical implementations frequently differ due to variations in design choices and auxiliary configuration parameters. For example, as illustrated in Figure~\ref{fig:matchingrules}, the \texttt{\small argsort} function in PyTorch and TensorFlow includes multiple control-related parameters such as \texttt{\small descending}, \texttt{\small stable}, and \texttt{\small direction}, whereas the same API in Keras, Chainer, and JAX accepts only two input functional parameters. Despite this discrepancy, the core computational logic of the function, sorting tensor elements along a given axis, remains the same.
To enable meaningful comparison of core functionality, it is therefore necessary to filter out such auxiliary parameters, which we refer to as control-related parameters. These auxiliary parameters generally do not affect the core computational semantics of the API. Instead, they control secondary aspects such as naming conventions, device placement, data types, reproducibility, logging, or training behaviors. Since these parameters typically have default initial values, we filter them out to focus exclusively on functional equivalence.
We construct a control parameter filter list based on a comparative review of official documentation in all five frameworks. Table~\ref{tab:control-params} summarizes the representative categories and keywords used for filtering.

\begin{table*}[t!]
\centering
\small
\caption{Common Categories and Examples of Control-Related API Parameters.}
\label{tab:control-params}
\begin{tabularx}{\linewidth}{>{\bfseries}l X}
\toprule
Category & Example Keywords \\
\midrule
Naming & \texttt{\small name}, \texttt{\small layer\_name}, \texttt{\small op\_name} \\
Device and Execution Control & \texttt{\small device}, \texttt{\small jit}, \texttt{\small compile}, \texttt{\small layout}, \texttt{\small autocast} \\
Data Type and Shape Configuration & \texttt{\small dtype}, \texttt{\small output\_dtype}, \texttt{\small input\_shape}, \texttt{\small output\_shape}, \texttt{\small validate\_shape} \\
Stability and Reproducibility & \texttt{\small seed}, \texttt{\small stable}, \texttt{\small deterministic} \\
Logging and Debugging & \texttt{\small verbose}, \texttt{\small debug}, \texttt{\small log\_level} \\
Sorting and Direction Control & \texttt{\small direction}, \texttt{\small ascending}, \texttt{\small descending} \\
\bottomrule
\end{tabularx}
\end{table*}

After filtering out control-related parameters, we compare the functional signatures of each API pair by analyzing the number of remaining functional parameters and their corresponding types. We compute parameter similarity based on two dimensions: \textit{Count Similarity} and \textit{Type Similarity}.

\textit{Count Similarity} measures the degree to which the two APIs define a comparable number of functional parameters. Let $n_1$ and $n_2$ denote the number of functional parameters in API A and API B, respectively. The count similarity is defined as:

\begin{equation}
{SIM_{Count}} = 1 - \frac{|n_1 - n_2|}{\max(n_1, n_2)}
\label{eq:count-sim}
\end{equation}

\textit{Type Similarity} is assessed by defining a unified abstract type space (e.g., \texttt{\small Tensor}, \texttt{\small int}, \texttt{\small bool}) that enables type-level comparison across frameworks. To facilitate cross-framework comparisons, we map framework-specific parameter types to a unified abstract type space, as shown in Table~\ref{tab:type-mapping}. For instance, PyTorch’s \texttt{\small Tensor}, TensorFlow’s \texttt{\small Tensor}, and JAX’s \texttt{\small Array} are all abstracted as \texttt{\small Tensor}. For each matched parameter pair, we assign a type compatibility score $S_i$, where $S_i = 1$ if the types are identical or deemed compatible, and $S_i = 0$ if no valid match is found.
The overall type similarity is the normalized average of matched scores:

\begin{table}[t!]
\centering
\small
\caption{Example of Equivalent Parameter Types Mapping.}
\label{tab:type-mapping}
\begin{tabularx}{\linewidth}{>{\bfseries}l *{5}{>{\centering\arraybackslash}X}}
\toprule
Type & PyTorch & TensorFlow & Keras & Chainer & JAX \\
\midrule
\texttt{\small Tensor} & \footnotesize{\texttt{\footnotesize Tensor}} & \footnotesize{\texttt{\footnotesize Tensor}} & \footnotesize{\texttt{\footnotesize Tensor}} & \footnotesize{\texttt{\footnotesize Variable}} & \footnotesize{\texttt{\footnotesize Array}} \\
\texttt{\small Shape} & \footnotesize{\texttt{\footnotesize List}} & \footnotesize{\texttt{\footnotesize Tuple}} & \footnotesize{\texttt{\footnotesize Tuple}} &\footnotesize{ \texttt{\footnotesize Sequence}} & \footnotesize{\texttt{\footnotesize List}} \\
\bottomrule
\end{tabularx}
\end{table}

\begin{equation}
{SIM_{Type}} = \frac{\sum_{i=1}^{k} S_i}{\max(n_1, n_2)}
\label{eq:type-sim}
\end{equation}

We compute the final similarity score as a weighted combination of count and type similarities:

\begin{equation}
\label{eq:simpara}
SIM_{Param} =  {SIM_{Count}} + {SIM_{Type}}
\end{equation}

As shown in Figure~\ref{fig:matchingrules}, \texttt{\small argsort} across all five frameworks are aligned to a normalized signature \texttt{\small (Tensor, int) $\rightarrow$ Tensor}, yielding both count and type similarity scores of 1. Differences in auxiliary parameters, e.g., \texttt{\small descending}, \texttt{\small stable}, are excluded during preprocessing. This example demonstrates that filtering control-related parameters and applying equivalent parameter type mappings can effectively unify structurally different yet functionally equivalent APIs.

These matching rules collectively form a unified basis for evaluating API equivalence. Each rule captures a complementary aspect, ranging from naming consistency and semantic alignment to structural compatibility. In the following section, we leverage these matching rules to systematically identify and select the best-matched API candidates.

\subsection{Matching APIs}
\label{sec:3.2}
\subsubsection{Matching Functionally Equivalent APIs}

To match functionally equivalent APIs, we adopt a multi-stage strategy that sequentially applies API name, description, and parameter similarity. APIs passing all stages are grouped into functionally equivalent API groups, which form the basis for input generation and differential testing.

\textit{Stage 1: Candidate Retrieval via API Name Similarity.}  
We compute the normalized Levenshtein distance-based name similarity $SIM_{APIName}$ (Eq.~\ref{eq:api-name-sim}) between the reference API and all candidates in the target framework. Candidates with $SIM_{APIName} \geq 0.5$ are retained for further evaluation, while those below this threshold are discarded. The 0.5 cutoff is widely used in approximate string matching~\cite{cohen2003comparison, euzenat2013ontology}, effectively filtering out unrelated APIs.
Besides candidates with identical or near-identical names, we retain those with partial lexical overlap~(e.g., \texttt{\small softmax\_cross\_entropy} vs. \texttt{\small crossentropy\_loss}) to avoid excluding functionally similar APIs due to naming differences. All remaining candidates proceed to semantic evaluation via description similarity.

\textit{Stage 2: Semantic Filtering via Description Similarity.}  
For each candidate retrieved in Stage 1, we compute its description similarity $SIM_{Des}$ (Eq.~\ref{eq:DS_cosine}) with the reference API using Sentence-BERT embeddings and cosine similarity. We then rank all candidates based on their similarity scores. To identify strong semantic matches, we retain the top-ranked candidates that show a clear similarity advantage over others in the list. Specifically, if the top candidate exhibits a large relative margin compared to the following one, it is treated as the best match. Otherwise, the top-3 candidates are preserved for structural verification. This ranking-based strategy avoids reliance on fixed thresholds and naturally adapts to the semantic closeness observed in the candidate pool.

\textit{Stage 3: Structural Verification via Parameter Similarity.}  
For each remaining candidate, we apply our parameter-based structural verification. Control-related parameters are removed based on a predefined list summarized in Table~\ref{tab:control-params}. We then compute parameter similarity $SIM_{Param}$ (Eq.~\ref{eq:simpara}), which captures both the number and type compatibility of functional parameters. A match is confirmed if $SIM_{Param} = 2$, indicating exact structural equivalence. If $SIM_{Param} < 2$, it suggests a mismatch in the functional input parameters, which may result in divergent behavior or altered semantics. We discard these candidates to ensure strict functional equivalence, maintain input consistency, and avoid potential false positives in subsequent fuzz testing.

This pipeline preserves the explainability and extensibility of rule-based approaches while leveraging semantic embedding and structural alignment to enhance reliability.

\subsubsection{Aligning Input Parameters}
\label{sec:aligning-parameters}

Once functionally equivalent APIs are matched across frameworks, it is critical to align their input parameters to ensure consistency. Differences in parameter names, ordering, and type annotations, even for logically identical APIs, can lead to mismatched inputs or unintended behavior. We adopt an alignment strategy: 

\textit{Type Filtering.}
We first eliminate control-related parameters by referencing the predefined control-parameter list presented in Table~\ref{tab:control-params}. This filtering process is automated by scanning parameter names against predefined keywords, ensuring only parameters influencing core computational functionality remain. Retained parameters are subsequently mapped to a unified type space (see Table~\ref{tab:type-mapping}), enabling precise cross-framework comparisons.

\textit{Name Normalization and Positional Alignment.}
To address inconsistencies in naming conventions, parameter names are first normalized to a canonical form using a manually curated alias mapping. Examples of common normalizations include:
\begin{itemize}[leftmargin=*]
\item \texttt{\small dim}, \texttt{\small axis}, \texttt{\small axes} $\rightarrow$ \texttt{\small axis}
\item \texttt{\small x}, \texttt{\small values}, \texttt{\small input} $\rightarrow$ \texttt{\small input}
\end{itemize}
Indexed parameters such as \texttt{\small tensor1}, \texttt{\small tensor2} are consolidated under a unified naming scheme.

We designate the parameter ordering of PyTorch APIs as the canonical signature. After name normalization, candidate API parameters from other frameworks are aligned to this canonical order by matching normalized parameter names, ensuring compatibility in parameter types, and resolving ambiguity by prioritizing exact name correspondence. If ambiguity persists, such as when multiple parameters share the same abstract type, positional proximity in the original signature is used as a secondary matching criterion. 
For instance, \texttt{\small torch.matmul(input, other)} aligns with \texttt{\small jax.numpy.matmul(a, b)} by matching \texttt{\small input} to \texttt{\small a} and \texttt{\small other} to \texttt{\small b}, reflecting both semantic intent and positional correspondence.

Through these steps, this strategy helps to match APIs from different frameworks to receive consistent input.  

\subsection{Fuzzing}
\label{sec:3.3}
\subsubsection{Input Seed Generation}

\tool\ generates input seeds for each matched API group using a unified parameter format:

\begin{itemize}[leftmargin=*]
\item \textit{Tensor}: Synthetic tensors are generated as multi-dimensional arrays using normal distributions. The shapes are randomly sampled from valid dimensions (e.g., 1D to 4D) to simulate realistic input scenarios, and tensors are cast into framework-native types (e.g., \texttt{\small torch.Tensor}, \texttt{\small tf.Tensor}, \texttt{\small jax.Array}) with numerical precision float32 or float64. To assess numerical robustness, edge-case values (\texttt{\small NaN}, \texttt{\small Inf}, extremely small or large floats) are also injected.

\item \textit{Value Scalar}: Scalar parameters representing tunable hyperparameters are generated as floating-point numbers uniformly sampled within a bounded range from $[0.0, 1.0]$, to prevent instability due to overflow or underflow.

\item \textit{Index Scalar}: Parameters that serve as dimension indices are selected based on the rank of the associated tensor, ensuring valid and meaningful indexing across frameworks.

\item \textit{Shape}: Shape-related parameters are generated as integer lists or tuples that align with the dimensionality of the associated tensors, ensuring validity and compatibility across frameworks.

\item \textit{Boolean Flag}: Boolean inputs are evaluated with both \texttt{\small True} and \texttt{\small False} to observe behavioral differences.
\end{itemize}
To enrich the diversity of generated inputs, we apply combinatorial testing principles by systematically varying parameter combinations, particularly for APIs with multiple arguments. For example, for the \texttt{\small argsort} API, we test both stable and unstable sorting modes with different axis values, and include edge cases such as empty tensors and tensors with repeated elements.
All inputs are first converted into a unified intermediate representation (e.g., NumPy arrays and Python native types) and then translated into each framework’s format. This representation helps mitigate inconsistencies introduced by framework-specific tensor initialization mechanisms, promoting comparable test conditions across APIs.

\subsubsection{Variance-Guided Differential Fuzzing}
\label{sec:Fuzzing}

After aligning matched APIs and generating consistent input seeds, we apply variance-guided differential testing to uncover behavioral inconsistencies. The core idea is to iteratively mutate inputs in a way that increases output divergence among functionally equivalent APIs.
Given a matched group of APIs \(\{A_1, A_2, \dots, A_n\}\), each API is treated as a black-box function with $m$ input arguments. Let \(\mathbf{x} = (x^{(1)}, x^{(2)}, \dots, x^{(m)})\) denote a complete input tuple, where each \(x^{(j)}\) satisfies the type, shape, and semantic constraints of the corresponding parameter. These constraints are derived from API signatures and normalized during the alignment phase.

Our goal is to identify valid inputs \(\mathbf{x}\) that trigger maximal behavioral divergence across the APIs. For each such input, we execute all APIs and observe their outputs.

To compute output variance $\sigma^2(\mathbf{x})$, we first standardize each API’s output by flattening multi-dimensional tensors into vectors and then compute element-wise variance. Specifically, we calculate the mean across APIs and the variance as:

\begin{equation}
\mu(\mathbf{x}) = \frac{1}{n} \sum_{i=1}^n A_i(\mathbf{x}),\quad
\sigma^2(\mathbf{x}) = \frac{1}{n} \sum_{i=1}^n \left(A_i(\mathbf{x}) - \mu(\mathbf{x})\right)^2
\end{equation}

where \(A_i(\mathbf{x}) \text{ is output of API } A_i \text{ on input } \mathbf{x}\), $\sigma^2(\mathbf{x})$ serves as a quantitative proxy for behavioral inconsistency. A higher variance indicates greater divergence among the outputs of semantically equivalent APIs.

To guide input mutation, we construct a deviation vector:
\begin{equation}
\mathbf{d}(\mathbf{x}) = \left[ A_1(\mathbf{x}) - \mu(\mathbf{x}), \dots, A_n(\mathbf{x}) - \mu(\mathbf{x}) \right],
\end{equation}
which indicates the direction in output space where APIs differ most significantly. We employ gradient-free perturbation strategies by adding scaled random noise aligned with $\mathbf{d}(\mathbf{x})$, systematically toggling boolean flags, and adjusting scalar inputs, to generate a mutated input $\mathbf{x}'$. Each perturbation maintains the shape, type, and semantic validity of the input parameters.
Our goal is to generate a mutated input $\mathbf{x}'$ that increases the output variance.
If the new input increases variance, it is accepted; otherwise, it is accepted with a small probability (simulated annealing) to escape local optima. If no improvement is observed after a predefined number of iterations, we reinitialize the input seed to escape stagnant regions of the input space and resume exploration. This feedback-driven mutation process iteratively explores high-divergence regions of the input space, allowing for the discovery of discrepancies under black-box assumptions.

\subsection{Oracles}
\label{sec:3.4}

In this section, we define the oracles to detect abnormal behaviors and inconsistencies during the execution of functionally equivalent deep learning APIs across frameworks.

\subsubsection{Crash} 
This oracle detects unexpected crashes during API execution. These include aborts, segmentation faults, assertion failures, and memory violations. Crashes typically indicate low-level bugs or implementation defects. If an API from one framework crashes while others execute successfully under the same input, the test case is marked as exposing a framework-specific vulnerability.

\subsubsection{NaN} 
This oracle checks whether any API returns a \texttt{\small NaN} (Not-a-Number) value during computation. \texttt{\small NaN} values usually arise from invalid mathematical operations (e.g., division by zero, numerical overflows). An input is flagged if only a subset of the APIs return \texttt{\small NaN}, indicating inconsistent numerical reliability across frameworks. Such discrepancies may silently compromise model correctness or downstream behavior.

\subsubsection{Inconsistency}
This oracle detects behavioral inconsistencies among APIs that are expected to be functionally equivalent. For a given input, we collect the outputs from all matched APIs and compute their variance using the metric \(\sigma^2(x)\) in Section~\ref{sec:Fuzzing}. A high variance implies that the APIs produce noticeably different results under the same input. 

These three oracles allow \tool\ to comprehensively test for crashes, numerical instability, and inconsistency, enabling deeper insight into the reliability of deep learning APIs.

\section{{Evaluation}}

\subsection{Research Questions}

We aim to investigate the following research questions:

\begin{itemize}
  \item[\textbf{RQ1:}] {What is the effectiveness of cross-framework matching for functionally equivalent APIs?}
  \item[\textbf{RQ2:}] {How do the fuzzing configurations \tool\ affect its effectiveness?}
  \item[\textbf{RQ3:}] How does \tool\ compare with existing methods?
  \item[\textbf{RQ4:}] {How effective is \tool\ in detecting bugs?}
\end{itemize}

For RQ1, we evaluate the effectiveness of our similarity-based matching rules in identifying functionally equivalent APIs across deep learning frameworks. We analyze the number and distribution of matched API groups. To evaluate accuracy, we test all matched API groups by executing them with aligned inputs and ensuring they produce consistent outputs across frameworks.
For RQ2, we evaluate how \tool's configuration influences its testing effectiveness. We examine two factors: (1) the number of input generation iterations and its effect on line-level code coverage, and (2) the fuzzing strategy by comparing our variance-guided approach with a random input baseline under the same number of iterations.
For RQ3, we evaluate how \tool\ compares with existing testing approaches for deep learning libraries in terms of code coverage and bug detection. 
For RQ4, we investigate the effectiveness of \tool\ in detecting inconsistencies across frameworks through differential testing. We report the number and types of identified inconsistencies, including crashes, \texttt{\small NaN} values, and high output differences. We also present case studies of real bugs discovered in popular frameworks to demonstrate the practical impact of \tool.

\subsection{Experimental Setup}

\noindent\textbf{Targeted Deep learning libraries.} 
This study evaluates five widely used deep learning frameworks, PyTorch, TensorFlow, Keras, Chainer, and JAX, covering diverse computational paradigms: static (TensorFlow), dynamic (PyTorch, Chainer), functional (JAX), and declarative high-level (Keras). These frameworks are widely adopted in academic and industrial settings, making them ideal for assessing the reliability and consistency of deep learning infrastructure.
We designate PyTorch as the reference baseline due to its mature ecosystem and comprehensive documentation, comparing the behavior of functionally equivalent APIs in other frameworks against it under unified input conditions.

All frameworks are installed within a shared Python environment (Python 3.9.20) with consistent dependencies, including NumPy (1.26.4) and SciPy (1.13.1). The framework versions evaluated are: PyTorch 2.2.1, TensorFlow 2.16.1, Keras 3.6.0, Chainer 7.8.1, and JAX 0.4.26. 
{We selected these versions because they can be installed together in a single environment without causing configuration conflicts.}
All APIs are invoked with logically equivalent inputs, first serialized into a unified intermediate form and then converted to framework-specific tensor types. 
Although our methodology is framework-agnostic and extendable to libraries like PaddlePaddle and MXNet, we restrict our evaluation to the five frameworks above. Preliminary tests showed that PaddlePaddle and MXNet introduced severe dependency conflicts (e.g., with protobuf, jaxlib, TensorFlow 2.x), making them infeasible for co-installation. Focusing on interoperable frameworks ensures reproducibility without relying on containerized environments.

\noindent\textbf{Targeted APIs.} 
We focus on 554 APIs defined under the torch and torch.nn namespaces, covering commonly used tensor creation, neural network construction, mathematical operations, reductions, and activation functions. These APIs serve as a practical baseline for cross-framework matching for two reasons: they capture core deep learning functionalities that are commonly used in libraries, and foundational operations like these are more likely to have counterparts in other frameworks, enhancing broader compatibility. 

\noindent\textbf{Baselines.}
We considered several recent approaches but excluded them due to technical limitations. {TitanFuzz}~\cite{deng2023largetitan} could not be included because its large language model(LLM) backend is inaccessible, making it impossible to run on compatible framework versions. {FuzzGPT}~\cite{deng2023large} is not available. {TensorScope}~\cite{deng2023differential} depends on translation-based testing, but its released code is incomplete as it is not runnable. As a result, we selected {FreeFuzz}~\cite{wei2022free} and {DeepREL}~\cite{deng2022fuzzing} as baselines, both of which offer open-source implementations and support API-level fuzzing for PyTorch and TensorFlow. Since these tools only support these two frameworks, our comparison is limited to PyTorch and TensorFlow for fairness.

\noindent\textbf{Testing budget.} 
For the experiment, we use \tool to generate 500 test cases per API, and evaluate the efficacy of the test of \tool. 

\noindent\textbf{Hyperparameter.} 
{During the matching stage, we set the relative margin threshold in \textit{Stage 2} to 0.3 and keep all the top-3 candidates whenever the margin falls below this threshold.}
During variance-guided differential fuzzing, input mutations without improvement are limited to 20 consecutive trials before reinitialization occurs. We empirically determine an acceptance threshold: a new input is considered improved only if it increases output variance $\sigma^2(x)$ by at least 0.001, balancing exploration and computational efficiency. An output variance threshold of $\sigma^2(x) \geq 0.1$ is selected based on preliminary empirical analyses, effectively distinguishing genuine behavioral inconsistencies while avoiding excessive false positives.

\noindent\textbf{Environment.} 
We use a 64-core PC with 96GB RAM and run on an NVIDIA RTX 6000ADA GPU.

\subsection{Metrics}

\noindent\textbf{Matched APIs.}
We provide statistics on the number of matched API groups and the total number of APIs involved, demonstrating the effectiveness of our similarity-based matching rules.

\noindent\textbf{Code Coverage.} 
We measure code coverage using both Python-level line coverage ({Coverage.py}~\cite{coveragepy}) and native backend C++ code coverage (\texttt{GCOV}~\cite{gcovmanual}). Coverage.py captures execution paths in Python-level APIs, whereas GCOV ensures the inclusion of underlying C++ implementations used by many deep learning operations. Employing both tools guarantees comprehensive coverage across all framework components, aligning with established practices in library fuzzing studies~\cite{gu2022muffin}.

\noindent\textbf{Detected bugs.} 
Following prior studies~\cite{xie2022docter, deng2023largetitan}, we count bugs identified as instances of three oracles. Each detected bug is reviewed to confirm reproducibility, thus ensuring the reliability of reported bugs.

\section{Result Analysis}
\subsection{RQ1: What is the effectiveness of cross-framework matching for functionally equivalent APIs?}

\subsubsection{Matched API Groups Across Frameworks}

We begin by examining the distribution of functionally equivalent API groups matched across the five frameworks.
In total, \tool\ identified \textit{\textbf{238 matched API groups}}, each containing semantically equivalent APIs from at least two frameworks. These groups collectively involve \textit{\textbf{839 individual APIs}} across all frameworks.
As shown in Table~\ref{tab:matched_group_stats}, {PyTorch} appears in all matched groups due to its role as the reference baseline, while {TensorFlow} and {Chainer} exhibit strong coverage with 226 and 200 matched APIs, respectively. In contrast, {Keras} and {JAX} have lower participation rates, which can be attributed to their relatively smaller API surface or higher-level abstraction layers. This distribution highlights the varying degrees of functional overlap between frameworks and the practical constraints in achieving uniform cross-framework equivalence.

\begin{figure}[t!]
  \centering
  \includegraphics[width=1\linewidth]{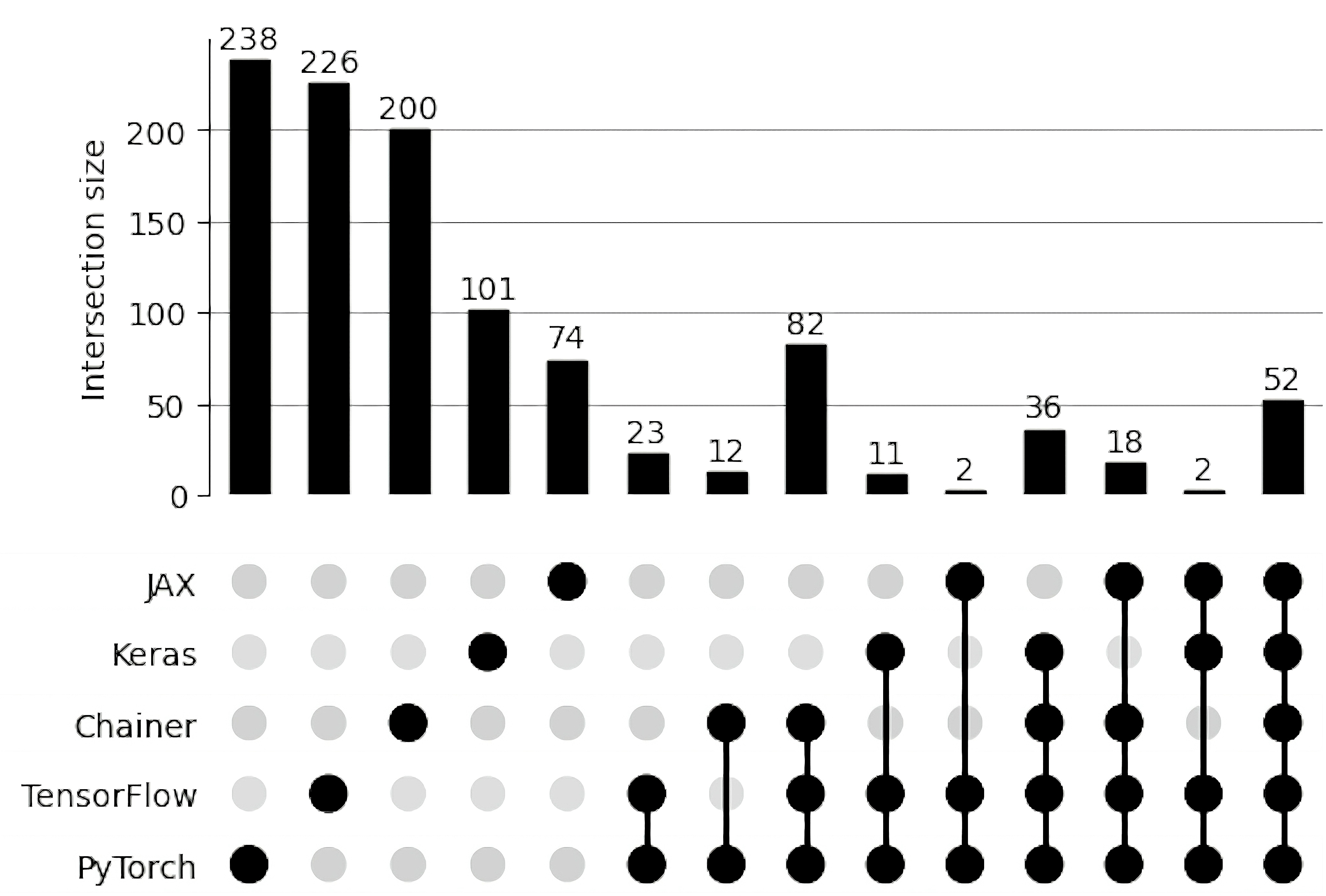}
  \caption{Functionally Equivalent API Group Matched.}
  \label{fig:api_upset_plot}
\end{figure}

To further analyze the structure and overlap of these matched API groups, we visualize their distribution across different framework combinations. 
Figure~\ref{fig:api_upset_plot} illustrates the distribution of matched API groups across different combinations of frameworks using an UpSet-style plot. Each bar at the top represents the size of the intersection, the number of functionally equivalent APIs, shared by the combination of frameworks indicated by the connected black dots below.
\begin{table}[t!]
\setlength{\tabcolsep}{4pt}
\centering
\small
\caption{Matched APIs Across Frameworks}
\label{tab:matched_group_stats}
\begin{tabular}{l|ccccc|@{\hskip 8pt}c}
\toprule
 & \textbf{PyTorch} & \textbf{TensorFlow} & \textbf{Keras} & \textbf{Chainer} & \textbf{JAX} & \textbf{Total} \\
\midrule
\#APIs & 238 & 226 & 101 & 200 & 74 & 839\\
\bottomrule
\end{tabular}
\end{table}

The leftmost five bars show the total number of matched APIs involving each individual framework: {PyTorch} (238), {TensorFlow} (226), {Chainer} (200), {Keras} (101), and {JAX} (74). These values confirm that {PyTorch} is involved in all 238 API groups, serving as the reference baseline in our matching process.
The bars from the sixth represent intersections across two or more frameworks. For example, there are 82 API groups that are matched across three frameworks, {PyTorch}, {TensorFlow}, and {Chainer}, while 52 API groups are matched across all five frameworks. 

Notably, to increase the confidence of functional equivalence, our design favors API groups that involve multiple frameworks. This makes it easier to cross-validate the consistency of behavior across multiple implementations. In practice, this goal is largely achieved: only 35 of the 238 matched API groups contain two APIs, while the remaining 203 groups involve three or more frameworks. This high degree of multi-framework alignment supports the reliability of the matched results and downstream differential testing.

\subsubsection{Behavioral Verification of Matched APIs}

To validate the correctness of the matched API groups produced by \tool, we conducted behavior-based verification. For each group of functionally equivalent APIs, we generated semantically consistent input seeds using the unified parameter alignment strategy and executed all matched APIs with the same inputs across frameworks.
We then compared the outputs using numerical equivalence checks. If all matched APIs produced consistent results on multiple test cases, we considered the match to be functionally valid.

\textbf{\textit{Behavioral validation showed all 238 matched API groups consistently produced highly similar outputs across 10 randomly generated input samples.}} The maximum observed difference, calculated as the absolute deviation by element, was less than 0.001, confirming functional equivalence at a numerical precision level sufficient for practical use.
This behavior-driven validation demonstrates the reliability of our similarity-based matching approach and provides evidence that matched APIs not only share structural similarities but also exhibit similar output across frameworks.

\subsection{RQ2: How do the fuzzing configurations \tool\ affect its effectiveness?}

We investigate how the following two key configurations influence \tool's effectiveness.

\subsubsection{Generation Times on Testing Effectiveness}

\begin{figure}[t!]
  \centering
  \includegraphics[width=1\linewidth]{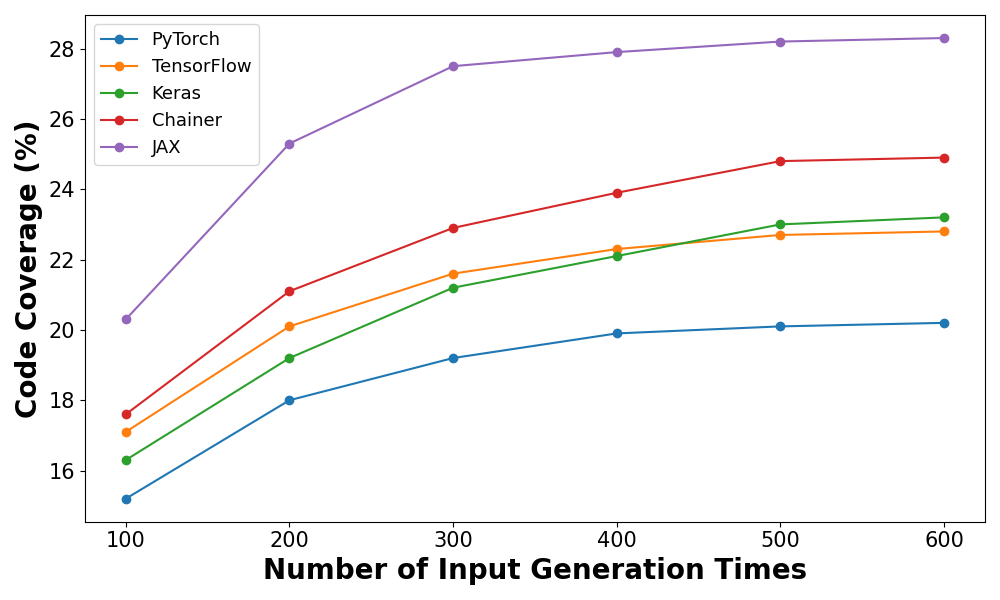}
  \caption{Code Coverage Trend Analysis Across Frameworks.}
  \label{fig:coverage}
\end{figure}

We demonstrate the effectiveness of \tool\ by analyzing how varying the number of generated test cases impacts code coverage across five deep learning frameworks: PyTorch, TensorFlow, Keras, Chainer, and JAX. As illustrated in Figure~\ref{fig:coverage}, the x-axis indicates the number of test cases generated per API (ranging from 100 to 600), while the y-axis represents the percentage of line coverage.

JAX starts with the highest initial coverage and shows steady growth before plateauing. PyTorch follows a similar gradual growth pattern, though with a lower initial coverage. TensorFlow, Keras, and Chainer exhibit comparable growth trends, initially increasing at a steady rate, but stabilizing around 500 test cases.

Overall, \textbf{\textit{the coverage growth for all frameworks begins to stabilize after approximately 500 test cases per API}}, suggesting this number provides a cost-effective balance between test case generation efforts and coverage gains. Therefore, we adopt 500 test cases as the optimal configuration.

\subsubsection{Comparison with Random Input Generation Strategy}

\begin{figure}[t!]
  \centering
  \includegraphics[width=1\linewidth]{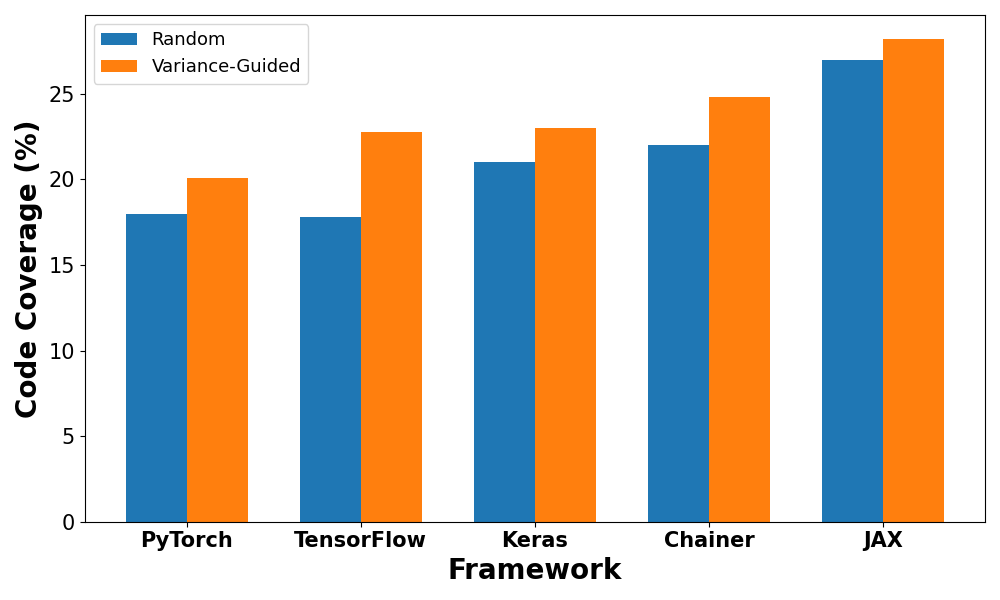}
  \caption{Compare with Variance-Guided and Random Strategy.}
  \label{fig:coveragecompare}
\end{figure}

To evaluate the impact of test generation strategies on code coverage, we compare \tool's variance-guided strategy with a purely random generation baseline, both configured to generate 500 test cases per API. The results are shown in Figure~\ref{fig:coveragecompare}, where each bar pair represents the coverage achieved on a specific framework.

Compared to random input generation, \tool's variance-guided strategy consistently achieved higher line coverage across all frameworks. Specifically, the feedback-driven mutations of inputs based on observed behavioral variance enabled deeper exploration of execution paths, significantly outperforming the random baseline.

For bug detection, we compared the baseline with our variance-guided strategy described in Section~\ref{sec:3.3}. \textbf{\textit{Under the same budget of generating 500 test cases per API group, the baseline was able to uncover only 9 bugs. In contrast, our divergence-amplified testing strategy successfully detected 17 bugs}}. This demonstrates that guiding input mutation towards maximizing behavioral divergence improves bug detection effectiveness compared to random input generation.

\subsection{RQ3: How does \tool\ compare with existing methods?}

\subsubsection{Code Coverage}
We compare \tool\ with two fuzzing-based testing methods, FreeFuzz, and DeepREL, on their ability to explore the execution space of deep learning frameworks. Since all baselines are designed specifically for PyTorch and TensorFlow, we restrict our comparison to these two frameworks to ensure a fair and consistent evaluation. This ensures that differences in framework compatibility do not bias the results.
Table~\ref{tab:coverage_sota} presents the code coverage achieved by each method. \textbf{\textit{In both PyTorch and TensorFlow, \tool\ consistently outperforms all baselines in terms of code coverage}}. This improvement stems from our differential testing input mutation strategy, which adaptively explores semantically meaningful paths based on behavioral divergence, enabling deeper and more diverse execution.

\begin{table}[t!]
\centering
\small
\caption{Comparison with Existing Methods.}
\label{tab:coverage_sota}
\resizebox{0.495\textwidth}{!}{
\begin{tabular}{lcc|cc}
\toprule
 & \multicolumn{2}{c|}{\textbf{PyTorch}} & \multicolumn{2}{c}{\textbf{TensorFlow}} \\
\cmidrule(r){2-3} \cmidrule(r){4-5}
\textbf{Method} & Coverage & Bugs & Coverage & Bugs \\
\midrule
FreeFuzz~\cite{wei2022free}   & 13.82\% & 0 & 20.15\% & 1 \\
DeepREL~\cite{deng2022fuzzing} & 13.91\% & 0 & 21.65\% & 1 \\
\textbf{\tool}                & \textbf{20.01\%} & \textbf{3} & \textbf{22.80\%} & \textbf{5} \\
\bottomrule
\end{tabular}
}
\end{table}

\subsubsection{Bug Detection}

To further compare with existing approaches, we applied FreeFuzz and DeepREL to the 8 API cases in PyTorch and TensorFlow where \tool\ detected bugs. Both methods only identified one crash bug in TensorFlow, failing to detect the rest. Most of these bugs exhibit no noticeable output differences between CPU and GPU, making them undetectable by intra-framework testing that relies on backend-induced discrepancies. Other backend-based methods share this limitation, as they depend on hardware-level variation to expose bugs. In contrast, 
\textbf{\textit{\tool\ complements existing approaches by leveraging a cross-framework oracle to uncover bugs that remain hidden under prior methods.
}}

\subsection{RQ4: How effective is \tool\ in detecting bugs?}

As shown in Table~\ref{tab:bug_count_categorized}, a total of 17 bugs were detected across the five major deep learning frameworks. These include 2 crash-related bugs, 8 bugs involving unexpected {NaN} outputs, and 7 inconsistency bugs where functionally equivalent APIs produced divergent results. Among them, {TensorFlow} and {JAX} exhibited the highest number of bugs (5), followed by {PyTorch} (3), {Keras} (2), and {Chainer} (2).

{Out of the 17 cases, 12 were confirmed by the developers of these frameworks, three inconsistencies were considered acceptable differences, and two cases (involving the Chainer framework) received no response.} All NaN bugs and most of the Inconsistency bugs were successfully confirmed. Through manual validation, we confirmed that no false positives among these.

The inconsistent output example in Listing~\ref{bug_nan_out} was triggered by applying the \texttt{\small argsort} API to an input array containing signed zeros, extremely small denormalized values, as well as large and negative floating-point numbers. This bug was caused by inconsistencies across frameworks when sorting elements with equal or nearly equal values. Previous testing methods that focused on identifying inconsistencies within different computational platforms could not detect such bugs, which can only be revealed through cross-framework comparison. We manually confirmed that modifying any single value in the array (e.g., replacing \texttt{\small -0.0} with \texttt{\small 0.0}) does not eliminate the inconsistency, indicating that the bug originates from the behavior of the internal sorting algorithm. This inconsistency has been reported and confirmed.

\begin{table}[t!]
\centering
\small
\caption{Bugs Detected and Confirmed}
\label{tab:bug_count_categorized}
\begin{tabular}{l c c c|c c}
\toprule
\textbf{Framework} & \textbf{Crash} & \textbf{NaN} & \textbf{Incon} & \textbf{Detected} & \textbf{Confirmed} \\
\midrule
PyTorch    & 0 & 1 & 2 & 3 & 2 \\
TensorFlow & 1 & 2 & 2 & 5 & 4 \\
Keras      & 0 & 1 & 1 & 2 & 2 \\
Chanier    & 0 & 1 & 1 & 2 & 0 \\
JAX        & 1 & 3 & 1 & 5 & 4 \\
\midrule
Total      & 2 & 8 & 7 & 17 & 12 \\
\bottomrule
\end{tabular}
\end{table}

\begin{figure}[t!]
\centering
\begin{minipage}{\linewidth}
\lstset{style=mystyle}
\begin{lstlisting}[language=Python, label=bug_nan]
input_data = np.array([ -0.0, 1.401298464324817e-45, 1.100000023841858, -0.0, 5.960464477539063e-08, -2.0000000135803223, 1000000.0, 722801.375, 0.0, -1.100000023841858], dtype=np.float32)
dim = 0
pytorch_result    = torch.argsort(input_data, dim)
tensorflow_result = tf.argsort(input_data, dim)
keras_result      = keras.argsort(input_data, dim)
chainer_result    = chainer.argsort(input_data, dim)
jax_result        = jax.argsort(input_data, dim)
\end{lstlisting}

\lstset{style=mystyle_1}
\begin{lstlisting}[language=Python, caption=Inconsistent Example 1, label=bug_nan_out]
output:
PyTorch result:     [5 9 0 3 8 1 4 2 7 6]
TensorFlow result:  [5 9 0 1 3 8 4 2 7 6]
Keras result:       [5 9 0 1 3 8 4 2 7 6]
Chainer result:     [5 9 0 1 3 8 4 2 7 6]
JAX result:         [5 9 0 1 3 8 4 2 7 6]
\end{lstlisting}
\end{minipage}
\label{fig:bug-nan-example}
\end{figure}

\begin{figure}[t!]
\centering
\begin{minipage}{\linewidth}
\lstset{style=mystyle}
\begin{lstlisting}[language=Python]
input_data = complex(float('nan'), float('nan')) 
pytorch_result    = torch.angle(input_data)
tensorflow_result = tf.math.angle(input_data)
jax_result        = jax.numpy.angle(input_data)
\end{lstlisting}


\lstset{style=mystyle_1}
\begin{lstlisting}[language=Python, caption=Inconsistent Example 2, label=bug_angle_nan_out]
output:
PyTorch result:     nan
TensorFlow result:  0.0
JAX result:         nan
\end{lstlisting}
\end{minipage}
\label{fig:angle-inconsistency}
\end{figure}

The inconsistent output example in Listing~\ref{bug_angle_nan_out} was triggered by setting both the real and imaginary parts of a complex number to {NaN}, and invoking the \texttt{\small angle} API. This input was designed to test how frameworks handle undefined or extreme behavior. The bug results from inconsistent \textit{NaN propagation semantics}: while {PyTorch} and {JAX} correctly propagate the {NaN} result, {TensorFlow} returns \texttt{\small 0.0}, silently masking the invalid input. We manually verified that modifying either the real or imaginary part to a finite value changes the output across frameworks, confirming that the inconsistency only occurs under specific {NaN} combinations. We reported this bug, and it has since been confirmed.

All of these bugs originate from flaws in the API implementations, even though these APIs are supposed to be functionally equivalent across different frameworks. \textit{\textbf{We verified that such NaN and Inconsistency bugs produce identical outputs across GPU and CPU, rendering them undetectable by previous methods that rely solely on backend-level discrepancies}}. This demonstrates the effectiveness of our approach in uncovering previously undetected bugs.

\section{Threats to Validity}

The primary internal threat to validity lies in the correctness of \tool’s matching algorithm implementation. To mitigate this, we conducted code-level validation to ensure that the similarity-based matching logic was accurately implemented as intended. We also performed targeted sanity checks to verify that key components, such as similarity computations and filtering stages, behaved as expected. 
{To mitigate cross-framework nondeterminism, we fix random seeds, disable nondeterministic kernels whenever possible, and repeat each experiment three times to compute the average results. During the matching stage, we retain control parameters that can affect numerical outcomes, such as {dtype} and {shape}, by normalizing them to a common precision or generating shapes from the abstract type space, ensuring semantic comparability across different frameworks.}

The external threat to validity concerns the generalizability of our findings across deep learning frameworks. Due to compatibility constraints with Python and dependency libraries, we included as many frameworks as possible. Ultimately, we selected five representative deep learning libraries and focused on testing their commonly used APIs.

\section{Related Work}

Deep learning has become integral to real-world applications, making the correctness of deep learning library APIs critical. Therefore, ensuring the correctness of APIs in these libraries is crucial. Researchers have applied various bug finding techniques to analyze them~\cite{yang2019synergistic,yang2019advances,yang2007algebraic,yi2022feedback,mahmud2022acid,yi2020summary}. One popular method is fuzzing~\cite{wu2022evaluating}, generating a large number of inputs and feeding them into the libraries to find potential bugs.
Existing approaches consist of fuzzing for deep learning frameworks\cite{christou2023ivysyn}, compilers\cite{xiao2022metamorphic}, computational graphs\cite{wang2022eagle}, and APIs\cite{deng2022fuzzing}. The focus of this paper is on fuzzing the functionality of deep learning APIs, which can be divided into two types: model-level fuzzing and API-level fuzzing.

Model-level fuzzing targets to generate a variety of complete models and then compare different model outputs to detect potential bugs. CRADLE~\cite{pham2019cradle} is one of the earliest works on model-level fuzzing, which detects inconsistencies by running existing models on Keras~\cite{chollet2015keras}. Then, to have more diverse models, LEMON~\cite{wang2020deep} and AUDEE~\cite{guo2020audee} further extended the idea of CRADLE by applying predefined mutation rules on the seed model and inputs. Recently, Muffin~\cite{gu2022muffin} applied a top-down approach to generate models for bug detection. NNSmith~\cite{liu2023nnsmith} utilizes symbolic constraint solving and gradient-based search to achieve high-quality model synthesis. Although existing model-level fuzzing techniques can find bugs, due to differences in inputs and outputs (such as type or size) of different APIs, model-level mutation and generation rules are mostly limited to a certain extent that maintains specific inputs and outputs, resulting in a limited number of APIs covered.
On the other hand, API-level fuzzing focuses on testing individual APIs by generating a variety of different inputs for each target API. FreeFuzz~\cite{wei2022free} is one of the earliest API-level fuzzing techniques, which learns valid inputs for each target API by mining open source code snippets and applying mutations to generate different inputs to test the target API. Similarly, DocTer~\cite{xie2022docter} mines input constraints from documentation by learning extraction rules for manually annotated API parameters, and then generates inputs based on the extracted constraints to detect crashes. DeepREL~\cite{deng2022fuzzing} further leverages relational APIs and automatic differentiation as test oracles to enable more efficient API-level fuzzing of deep learning libraries.  
However, these techniques can only detect bugs in individual APIs and are unable to test for bugs that occur in the interaction of multiple APIs.
Titanfuzz \cite{deng2023largetitan} uses LLM for fuzzing deep learning libraries and learns API constraints to facilitate deep learning computations, which use mathematical operations or type conversion APIs to modify the inputs or outputs of the target API and test the computational dependencies between APIs. 
FuzzGPT \cite{deng2023large} hypothesizes that historically bug-inducing programs may contain code components crucial to uncovering edge cases, which reconstruct and embed them in new code contexts to expose vulnerabilities \cite{chen2019history, holler2012fuzzing}. 

Despite these advances, these approaches primarily rely on differences arising from hardware backends (CPU vs. GPU). As a result, these methods typically miss bugs that affect implementations equally across different hardware configurations, such as algorithmic errors or numerical inconsistencies inherent to the framework's implementation logic.
TensorScope~\cite{deng2023differential} identifies converting APIs by mining the internal logic and parameter rules from model converters, focusing on API behaviors observed in model transformation scenarios, which reported bugs often stem from converter-induced errors (e.g., low-precision handling, mismatched parameters), not the native APIs themselves, making it more suitable for evaluating model portability than implementation correctness. 
Additionally, TensorScope’s reliance on model converters makes it limited to frameworks that have available converters. 
In contrast, our method directly uses official documentation to match APIs based on name, description, and parameter similarity, without relying on any external tools. By focusing on the native implementation of APIs across different libraries, our approach enables native and framework-agnostic testing that is not restricted by the availability or maintenance of model converters. As a result, our methodology can be adapted to a wider range of frameworks, providing high portability across libraries.

\section{Limitation}

Our design adopts three oracles, Crash, NaN, and Inconsistency, to detect abnormal behaviors, yet some incorrect behaviors may escape these checks. {The detected bugs indicate reliability issues that can cause silent failures and unstable model behavior, which are critical in safety-sensitive domains such as autonomous driving and facial recognition. A limitation of \tool\ is that common-mode bugs, which refer to identical flawed logic shared by multiple frameworks, may evade detection by cross-framework oracles; integrating metamorphic relations or high-precision reference checks is a promising direction to address such cases in future work. Furthermore, runtime scales approximately with the number of matched API groups and tests per API; \tool\ parallelizes across groups and can trade detection power for speed by tuning parallelism and the number of generated tests.}
Our evaluation currently targets APIs under the {torch} and {torch.nn} namespaces; expanding it to other namespaces or frameworks could improve the generalizability of the results.
Additionally, our strict API matching prevents incorrect matches and avoids manual intervention. However, it may also filter out some functionally equivalent APIs that differ in naming conventions or parameter designs. 
{A potential future enhancement is to leverage large language models to perform more accurate documentation extraction and matching, as demonstrated in VistaFuzz~\cite{duan2025harnessing}.}
Finally, our cross-framework testing complements existing intra-framework fuzzing approaches. While intra-framework methods effectively detect bugs exposed by CPU–GPU discrepancies, \tool\ targets bugs that produce consistent outputs across hardware but are incorrect. Existing methods enhance input diversity within frameworks, whereas \tool\ broadens the testing space across frameworks, making the two approaches complementary for improving deep learning library reliability.

\section{Conclusion}
{This paper presents \tool, a cross-framework fuzzing method that tests deep learning libraries by matching
and comparing functionally equivalent APIs across different
frameworks.} \tool\ identifies equivalent APIs across frameworks using similarity-based rules derived from documentation and generates consistent inputs for differential testing. By comparing outputs under shared inputs, \tool\ effectively uncovers inconsistencies and implementation bugs that are difficult to detect through intra-framework testing.
We evaluated \tool\ on five widely-used deep learning libraries, PyTorch, TensorFlow, Keras, Chainer, and JAX, covering 839 APIs and identifying 238 groups of functionally equivalent APIs. \tool\ detected 17 bugs, 12 of which have been confirmed, demonstrating its effectiveness in revealing bugs. These results highlight the potential of cross-framework fuzzing to improve the reliability of deep learning infrastructure.



\balance
\bibliographystyle{plain}
\bibliography{main}

\end{document}